# Integration of an optical fiber taper with an optical microresonator fabricated in glass by femtosecond laser 3D micromachining

Jiangxin Song[a,b], Jintian Lin[a,b], Jialei Tang[a,b], Lingling Qiao[a], Ya Cheng[a,*]

[a] *State Key Laboratory of High Field Laser Physics, Shanghai Institute of Optics and Fine Mechanics, Chinese Academy of Sciences, Shanghai 201800, P. R. China.*
[b] *Graduate School of Chinese Academics of Science, Beijing 100039, China*
[*] *Email: ya.cheng@siom.ac.cn*

## Abstract

We report on fabrication of a microtoroid resonator of a high-quality factor (i. e., Q-factor of ~$3.24\times10^6$ measured under the critical coupling condition) using femtosecond laser three-dimensional (3D) micromachining. Coupling of light into and out of the microresonator has been realized with a fiber taper that is reliably assembled with the microtoroid. The assembly of the fiber taper to the microtoroid is achieved by welding the fiber taper onto the sidewall of the microtoroid using $CO_2$ laser irradiation. The integrated microresonator maintains a high Q-factor of $3.21\times10^5$ as measured in air.

For many optofluidic sensing applications, efficient enhancement of the sensing sensitivity can be achieved by use of a high-quality factor (Q factor) optical cavity [1-3]. In particular, it has been shown recently that a whispering gallery mode (WGM) microresonator can be used for biosensing with a detection limit down to single molecules, owing to its unique capability of confining light in a tiny volume for long periods of time by total internal reflection [4,5]. The WGM microresonators demonstrated in these experiments are typically microspheres or microtoroids. In most cases, coupling of light into and out of the resonators is realized with a fiber taper carefully positioned near the microresonator with precision motion stages to achieve the critical coupling condition [6]. Since the fiber taper and the microresonators are not integrated, the widespread use of such systems in optofluidic and biosensing applications has been hampered because of their poor portbility. Further chip-level integration of the microresonators in a microfluidic system is also difficult, because of the incompatibility between the processing techniques of microfluidic systems and optical microresonators. In our opinion, these issues could potentially be resolved based on fabricating the microcomponents of different functions in a common substrate using femtosecond laser three-dimensional (3D) micromachining [7,8].

As we have demonstrated in 2012, WGM microresonators of Q-factors on the order of $10^6$ and of various 3D geometries can be fabricated in fused silica using femtosecond laser direct writing followed by chemical wet etching, i.e., the same approach that has been employed for fabrication of microfluidics in fused silica [9,10]. Herein, we further show that a fiber taper can be reliably assembled with the microresonator by $CO_2$ laser welding for coupling light into and out of the resonator. The fully integrated on-chip microresonator system can be used for label-free field-portable sensing applications.

In our experiment, commercially available fused silica glass substrates (UV grade fused silica JGS1 whose upper and bottom surfaces were polished to optical grade) with a thickness of 1 mm were used. The process flow for fabrication of the fully integrated sensor mainly consists of three steps: (1) femtosecond laser exposure followed by selective wet etch of the irradiated areas to create a 3D microfluidic mixer embedded in glass and a microdisk structure near the outlet of the microfluidic mixer; (2) selective reflow of the silica disk by $CO_2$ laser irradiation to improve the Q-factors; and (3) assembly of the fiber taper with the microresonator by $CO_2$ laser welding. The experimental parameters are described as follows.

The femtosecond laser system consists of a Ti: sapphire oscillator (Coherent, Inc.) and a regenerative amplifier, which emits 800 nm, ~40 fs pulses with maximum pulse energy up to of ~4.5 mJ at 1 kHz repetition rate. The initial 8.8-mm-diameter beam was trimmed to 5 mm-diameter by passing the beam through a circular aperture to improve the beam quality. Power adjustment was realized using a set of neutral density (ND) filters. The glass samples could be arbitrarily translated in 3D space with a resolution of 1 μm by a PC-controlled XYZ motion stage combined with a nano-positioning stage.

The high-Q microtoroid resonator was created by first forming a microdisk supported by a thin pillar in glass. The microdisk and the thin pillar were both fabricated in a single scan by use of a layer-by-layer annular scanning with the nano-positioning stage. The lateral scanning step was set at 1 μm, and the scanning speed was set at 600 μm/s. A 100× objective with a numerical aperture (NA) of 0.8 was used to focus the beam to a ~1 μm-dia. spot, and the average femtosecond laser power measured before the objective was ~0.05 mW. After the laser writing, the sample was etched for 0.5 hrs with hydrofluoric (HF) acid diluted in water at a volume concentration of 5%.

After the chemical wet etching, the sidewall of the fabricated microdisk had a roughness of several tens of nanometer, which is not suitable for high-Q resonator applications.

To improve the surface smoothness on the sidewall of the microdisk and achieve the desirable high Q-factor, we smoothed the surface of the microdisk by introducing a surface-normal-irradiation with a $CO_2$ laser (Synrad Firestar V30). The $CO_2$ laser beam was focused to a circular spot of approximately ~100 μm in diameter by a lens. The laser was operated with a repetition rate of 5 kHz, and the irradiation strength was controlled by adjusting the duty ratio. As the disk diameter thermally shrinked during the reflow, surface tension induced a collapse of the fused silica disk, leading to a toroid-shaped boundary. During this process the disk was monitored by a charge coupled device (CCD) from the side using a 20× objective lens. The total reflow process took merely 4 s at a duty ratio of 5.0%. Thanks to the reflow, the smoothness of the microtoroid surface is excellent, as shown in the optical micrograph [see, e.g., Fig. 1(b)]. The microtoroid was measured to have a diameter of ~80 μm and a ~10-μm-thick toroid-shaped boundary. More details for fabrication of the microtoroid resonator can be found in our previous publications [9,10].

To characterize the mode structure and the Q factor of the assembled microtoroid cavity, resonance spectra were recorded with the optical fiber taper coupling method. A swept-wavelength tunable external-cavity diode Laser (New Focus, model: 6528-LN) and a swept spectrometer (dBm Optics, model: 4650) were used to measure the transmission spectrum from the fiber taper with a resolution of 0.1 pm. The fiber taper was fabricated by heating and stretching a section of optical fiber (SMF-28, Corning) until reaching a minimum waist diameter of approximately ~1 μm. The fiber taper can provide an evanescent excitation of whispering gallery modes of the cavity. The position of the fiber taper was controlled using a three-axis nano-positioning stage with a spatial resolution of 50-nm in the XYZ directions. We used dual CCD cameras to simultaneously image the microtoroid cavity and fiber taper from both the side and the top [10].

The key innovation of this work is to assemble the fiber taper onto the sidewall of the microtoroid resonator to produce a fully integrated, field-portable microcavity unit. To this end, the fiber taper was brought to the vicinity of the toroid using the 3D nano-positioning stage until the critical coupling condition was realized. The highest Q-factor we obtained under the critical coupling condition was $3.24 \times 10^6$, as evidenced by the transmission spectrum provided in Fig. 2. Then we slightly moved the fiber taper to have it in direct contact with the sidewall of the microtoroid, and irradiated the microtoroid with the $CO_2$ laser beam again, as schematically illustrated in Fig. 1(a). We note that the focusing conditions of the $CO_2$ laser beam used for assembling the fiber were the same as that used for reflow the microdisk to form the microtoroid, namely, no modification was applied on the original optical setup which was established for the surface reflow of the microresonator. However, to achieve a high Q-factor of the assembled microtoroid-fiber system, indeed we had optimized the duty ratio of the $CO_2$ laser. For the welding, eventually we chose a duty ratio of 4.0% (i.e., 20% lower than that used for reflowing the microdisk), leading to both a strong welding and a decent Q-factor in air of $3.21 \times 10^5$ (see Fig. 3 below).

Figure 1(b) shows the optical micrograph of the microtoroid assembled with the fiber taper. One can notice that the $CO_2$ laser irradiation creates a strong welding in the contacting area between the fiber taper and the microtoroid, thus the fiber can be strongly bent without falling apart from the microtoroid. The high mechanical strength will ensure a reliable operation of the integrated microcavity system for field-portable uses.

Figure 2(a) shows the transmission spectrum of the femtosecond laser fabricated microtoroid measured under the critical coupling condition. The free spectral range of 5.66 nm agrees well with the theoretical calculation. As shown in Fig. 2(b), the Q-factor calculated based on a Lorentz fitting reaches $3.24\times10^6$, which is slightly better than our previous results [9,10].

For comparison, the transmission spectrum of the integrated microtoroid-fiber system is shown in Fig. 3(a). The spectrum shows more complicated structure mainly because of the excitation of high-order modes. Again, based on the Lorentz fitting (i.e., the red curve in Fig. 3(b)), the Q-factor is estimated to be $\sim3.21\times10^5$. This is about one order of magnitude lower than the Q-factor achieved under the critical coupling condition. However, for many sensing applications, the moderate sacrifice on the Q-factor is acceptable because the high detection limits can still be obtained at this Q level. Meanwhile, the portability of the sensors has been dramatically improved, which is highly desirable for field-portable applications.

At last, we present the overall transmission spectra of the integrated microcavity which were measured in both air (black curve) and water (red curve), as shown in Fig. 4. It can be seen that as compared to the transmission spectrum measured in air, the transmission spectrum measured in water shows larger spectral width of each dip and reduced number of modes, mainly because of the reduced Q-factor and the elimination of the high-order modes, respectively. The measured Q-factor of the microresonator immersed in purified water was evaluated to be $\sim3.1\times10^3$. However, the significant reduction of the Q-factor in water is avoidable for the following reasons.

We have clarified that the reduction of the Q in water is mainly caused by two factors. The first limiting factor is that the experiment was carried out at a wavelength of ~1550 nm (i.e., the only tunable laser source available in our lab), which is highly absorptive in water [11]. The second is that because of the limited range of motion of our nano-positioning stage, the size of the fabricated microtoroid resonator is limited to ~80 μm, which is not large enough to support sufficiently high Q-factors in water as have been investigated before. Actually, in Ref. [12], the authors have clearly shown that for a fused silica microtoroid with a diameter of ~80 μm, it exhibits a Q-factor on the level of $\sim 10^3$ in water in the spectral range around 1550 nm. The measured Q-factor of our fabricated resonator (Q~$3.1 \times 10^3$) agrees well with their analysis. Thus, by replacing the infrared excitation beam with a laser source in the visible range and increasing the diameter of the microtoroid resonator, the Q-factor of the fully integrated microcavity system can be efficiently promoted until reaching its up limit of ~$3.2 \times 10^5$, i.e., the Q-factor of the system in air.

In conclusion, we have demonstrated fabrication of a microresonator assembled with a fiber taper using femtosecond laser direct writing followed by $CO_2$ laser welding. The microtoroid resonator exhibits a Q-factor of $3.24 \times 10^6$ in air under the critical coupling condition. After the fiber taper is welded onto the sidewall of the resonator, the Q-factor of the integrated system measured in air decreases to ~$3.21 \times 10^5$. By incorporating the integrated microcavity system into a 3D microfluidic network which can also be realized in glass by femtosecond laser micromachining, field-portable optofluidic sensing can be performed with high detection limits thanks to the high-Q nature of the WGM microresonators.

**References:**

[1] W. Z. Song, X. M. Zhang, A. Q. Liu, C. S. Lim, P. H. Yap, and H. M. Hosseini, *Appl. Phys. Lett.*, 2006, **89**: 203901.

[2] F. Vollmer, D. Braun, A. Libchaber, M. Khoshsima, I. Teraoka, S. Arnold, *Appl. Phys. Lett.*, 2002, **80**, 4057-4059.

[3] A. M. Armani, R. P. Kulkarni, S. E. Fraser, R. C. Flagan, and K. J. Vahala, *Science*, 2007, **317**, 783-787.

[4] F. Vollmer and S. Arnold, *Nature Meth.*, 2008, **5**, 591-596.

[5] J. Zhu, S. K. Ozdemir, Y.-F. Xiao, L. Li, L. He, D.-R. Chen, and L. Yang, *Nature Photon.*, 2010, **4**, 46-49.

[6] F. Vollmer, and L. Yang, *Nanophoton.*, 2012, **1**, 267-291

[7] K. Sugioka and Y. Cheng, *Lab Chip*, 2012, **12**, 3576-3589.

[8] R. Osellame, H. J. W. M. Hoekstra, G. Cerullo1 and M. Pollnau, *Laser Photonics Rev.*, 2011, **5**, 442–463.

[9] J. Lin, S. Yu, Y. Ma, W. Fang, F. He, L. Qiao, L. Tong, Y. Cheng, and Z. Xu, *Opt. Express*, 2012, **20**, 10212-10217.

[10] J. Lin, S. Yu, J. Song, B. Zeng, F. He, H. Xu, K. Sugioka, W. Fang, and Y.Cheng, *Opt. Lett.*, 2013, **38**, 1485-1487

[11] G. M. Hale and M. R. Querry, *Appl. Opt.*, 1973, **12**, 555-563.

[12] A. M. Armani, D. K. Armani, B. Min, and K. J. Vahala, *Appl. Phys. Lett.,* 2005, **87**, 151118.

**Figure captions**

Fig. 1 (a) Schematic illustration of assembling a fiber taper onto the microresonator. (b) Side-view optical micrograph of the fiber taper welded onto the sidewall. Fiber can remain assembled on the microtoroid even it is severely bent.

Fig. 2 (a) Transmission spectrum of the microtoroid measured under the critical coupling condition. (b) The Rorentz fitting (red curve) of one dip in the spectrum, showing a Q-factor of $3.24 \times 10^6$.

Fig. 3 (a) Transmission spectrum of the integrated microcavity system measured in air after the $CO_2$ laser welding. (b) The Rorentz fitting (red curve) of one dip in the spectrum, showing a Q-factor of $3.21 \times 10^5$.

Fig. 4 Transmission spectrum of the integrated microcavity system measured in air after the $CO_2$ laser welding (black curve) and that of the integrated microcavity system measured in purified water (red curve).

Fig. 1

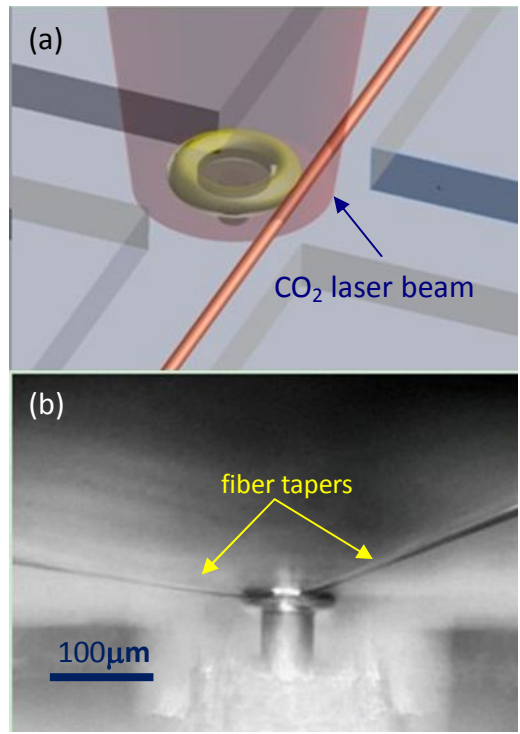

Fig. 2

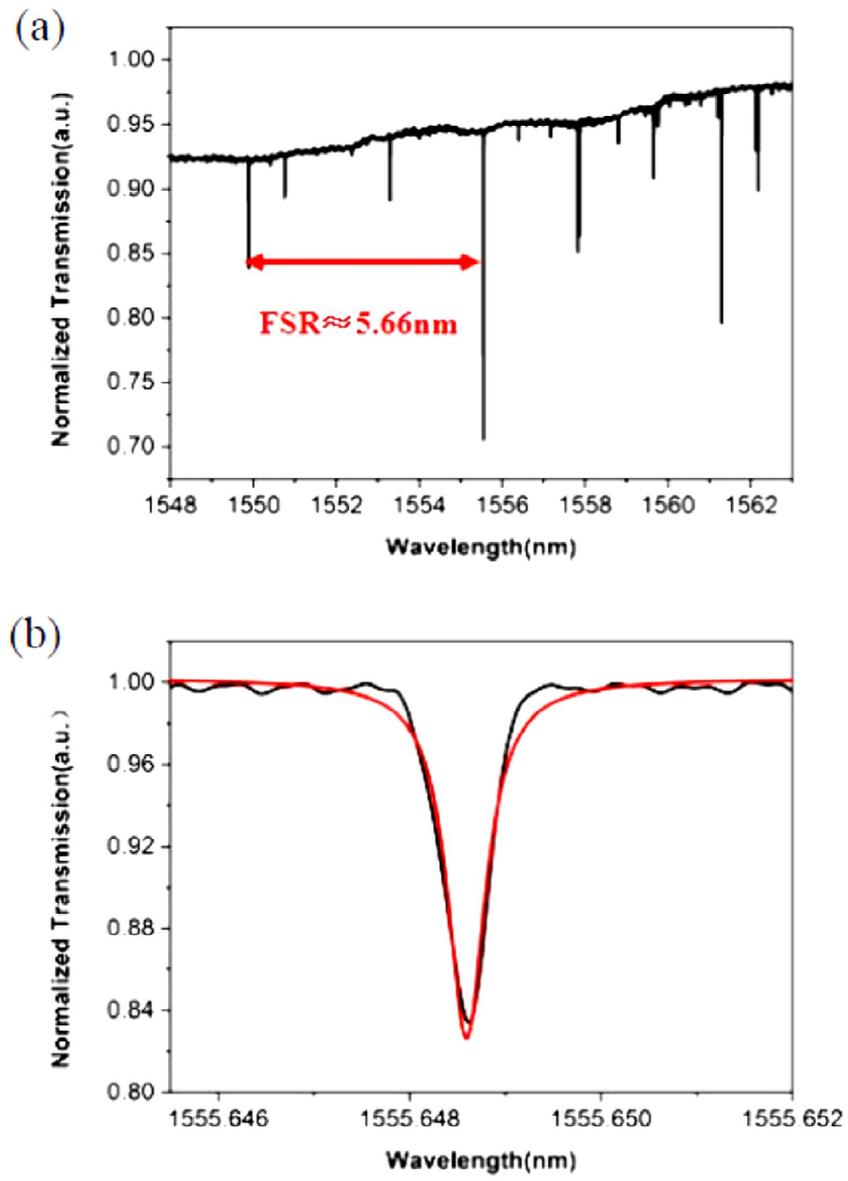

Fig. 3

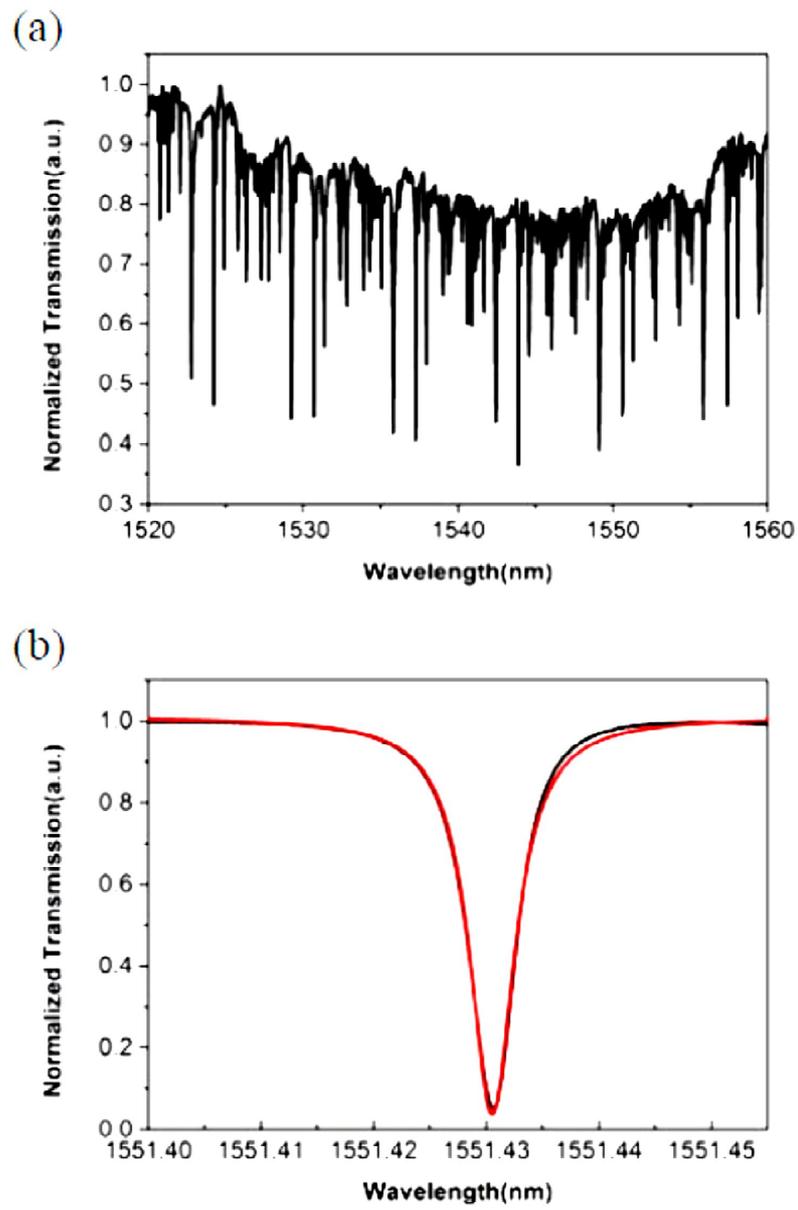

Fig. 4

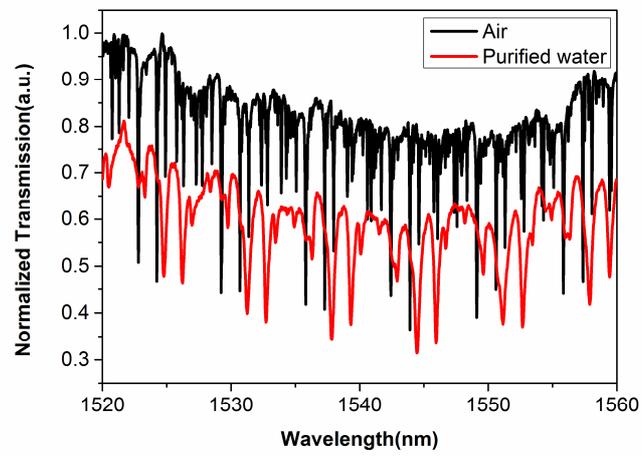